\documentclass[a4paper]{jpconf}

\usepackage{graphicx}
\usepackage{amsmath}
\usepackage{amscd}
\usepackage{amsfonts}
\usepackage{amssymb}

\begin{document}

\title{Trivial symmetries in a 3D topological torsion model of gravity}

\author{Rabin Banerjee$^{a,1}$ and Debraj Roy$^{a,2}$}

\address{$^a$SN Bose National Centre for Basic Sciences, Salt Lake, Blk JD, Sec 3, Kolkata - 700098.}

\ead{$^1$rabin@bose.res.in, $^2$debraj@bose.res.in}

\begin{abstract}
We study the gauge symmetries in a Mielke-Baekler type model of gravity in 2+1 dimensions. The model is built in a Poincare gauge theory framework where localisation of Poincare symmetries lead to gravity. However, explicit construction of gauge symmetries in the model through a Hamiltonian procedure yields an apparently different set of symmetries, as has been noted by various authors. Here, we show that the two sets of symmetries are actually equivalent in a canonical sense, their difference being just a set of trivial symmetries.
\end{abstract}

\section{Introduction}

Gauge symmetries and their systematic construction, while being of great significance, is fraught with subtleties. They must be off-shell, i.e. the action should remain invariant under them without use of any equations of motion and there are systematic algorithms for construction of such off-shell symmetries \cite{Henneaux:1990au,Banerjee:1999yc,Banerjee:1999hu}. In models of gravity obtained by gauging Poincare symmetries, it is however seen that such systematic approaches yield the Poincare symmetries only on-shell. So, at the off-shell level, it appears that there are two independent set of symmetries. This is unsettling. We would then be required to find additional independent first-class constraints to support these symmetries. However, physically the only independent symmetries present are the space-time translations and Lorentz boosts, i.e. the Poincare symmetries.

We set up the model in the Poincare gauge theory (PGT) formalism \cite{Utiyama:1956sy,Kibble:1961ba,Sciama:1962}. The global Poincare symmetries on a set of global coordinates $x^\mu$ are localised at every spacetime point by introducing local frames having coordinates $x^i$. Both sets are 3-dimensional for our model, though the construction can be carried out in any dimensions. To change physical variables from one set of coordinates to another, triad fields are introduced, which are defined through $$g_{\mu\nu}=b^i_{\ \mu}\,b^j_{\ \nu}\,\eta_{ij}\,.$$ Here `$g$' and `$\eta$' are the metrics of the global and local sets of coordinates, respectively and `$\eta$' is the standard Minkowskian metric with signature  $(+,-,-)$. Now, to keep the invariance of the theory, a gauge connection field `$\omega^i_{\ \mu}$' is introduced through a newly defined gauge covariant derivative. For an arbitrary vector $A^i$, the derivative is written as $$\nabla_\mu A^i = \partial_\mu A^i + \epsilon^i_{\ jk}\omega^j_{\ \mu}A^k\,.$$ The corresponding field strengths give rise to gravitational fields
\begin{align*}
\left[\nabla_\mu,\nabla_\nu\right] b^i_{\ \rho} = \epsilon^i_{\ jk} R^j_{\ \mu\nu} b^k_{\ \rho}\,,
\end{align*}
where $R^i_{\ \mu\nu} = \partial_\mu \omega^i_{\ \nu} - \partial_\mu \omega^i_{\ \nu} + \epsilon^i_{\ jk}\,\omega^j_{\ \mu}\omega^k_{\ \nu}$ defines the Riemann. Similarly, the commutator of $\nabla_i$ derivatives give rise to the torsion $T^i_{\ \mu\nu} = \nabla_\mu b^i_{\ \nu} - \nabla_\nu b^i_{\ \mu}.$ It must be emphasised that this is a setting in the Riemann-Cartan spacetime rather than the Riemann spacetime of usual general relativity. In general it contains torsion, which goes to zero only after imposition of the connection variable's equation of motion, giving back the torsion free Riemannian manifold of general relativity. The Poincare symmetries of the basic fields are \cite{Banerjee:2009vf}
\begin{align}
\label{PGT deltas}
\begin{aligned}
\delta_{\scriptscriptstyle PGT} b^i_{\ \mu} &= -\epsilon^i_{\ jk}b^j_{\ \mu}\theta^k - \partial_\mu \xi^\rho \,b^i_{\ \rho} - \xi^\rho\,\partial_\rho b^i_{\ \mu} \\
\delta_{\scriptscriptstyle PGT} \omega^i_{\ \mu} &= -\partial_\mu \theta^i - \epsilon^i_{\ jk}\omega^j_{\ \mu}\theta^k - \partial_\mu\xi^\rho\,\omega^i_{\ \rho} - \xi^\rho\,\partial_\rho\omega^i_{\ \mu}\,,
\end{aligned}
\end{align}
where $\theta^i$ parametrizes local Lorentz rotations and $\xi^\rho$ local coordinate translations. They together constitute the $3+3=6$ independent Poincare symmetries in 3D. These symmetries are independent of the particular action being considered, so long as they are constructed in a proper tensorial manner.

Now, symmetries of an action can be constructed via a Hamiltonian analysis through completely off-shell procedures, as stated before. After this is carried out, we find the symmetries which keeps the action invariant, as expected. However they cannot be identified with the Poincare symmetries \eqref{PGT deltas}, even after an appropriate field-dependent redefinition of the symmetry parameters. The clue of identification, as we show, lies in `trivial' gauge symmetries \cite{Henneaux:1992ig,Henneaux:2009zb}. They are symmetries of the form
\begin{align}
\label{trivial gauge gen}
\delta q_i = \Lambda_{ij}\,\frac{\delta S}{\delta q_j}\qquad \left(\Lambda_{ij}=-\Lambda_{ji}\right),
\end{align}
which keep any action {\it off-shell} invariant due to antisymmetry in the coefficients $\Lambda_{ij}$. Thus
\begin{align}
\label{var action for trivial}
\delta S &= \frac{\delta S}{\delta q_i} \delta q_i = \frac{\delta S}{\delta q_i}\,\Lambda_{ij}\,\frac{\delta S}{\delta q_j} = 0
\end{align}
without depending on the form of the Euler derivatives $\frac{\delta S}{\delta q_i}$ and without requiring the imposition of the equations of motion $\frac{\delta S}{\delta q_i} = 0\,.$

In what follows, we take up the Mielke-Baekler model, summarise its Hamiltonian constraints, outline construction of an off-shell gauge generator\footnote{Note that there exist methods \cite{Castellani:1981us} which construct symmetries of fields that keep the equations of motion invariant. We are using a different approach.} and generate gauge symmetries. These are then compared with the Poincare symmetries to see how the two may actually be equivalent.

\section{Hamiltonain analysis and construction of a gauge generator}

The Mielke-Baekler 3D topological gravity model with torsion \cite{Baekler:1992ab} along with a cosmological term is described by
\begin{align}
\label{action TMG}
S = \!\int \!\textrm{d$^3$x}\,\epsilon^{\mu\nu\rho}\!\left[ab^i_{\ \mu}R_{i\nu\rho} - \frac{\Lambda}{3} \epsilon_{ijk}b^i_{\ \mu}b^j_{\ \nu}b^k_{\ \rho} + \alpha_3\!\left(\!\omega^i_{\ \mu}\partial_\nu\omega_{i\rho}  + \frac{1}{3} \epsilon_{ijk}\,\omega^i_{\ \mu}\omega^j_{\ \nu}\omega^k_{\ \rho} \right) + \frac{\alpha_4}{2}b^i_{\ \mu}T_{i\nu\rho} \right]
\end{align}
whose equations of motion are found by setting the Euler derivatives to zero
\begin{align}
\label{EOM MB}
\begin{aligned}
\frac{\delta S}{\delta b^i_{\ \mu}} &= \epsilon^{\mu\nu\rho} \left[ a\,R_{i\nu\rho} + \alpha_4\, T_{i\nu\rho} - \Lambda\, \epsilon_{ijk}b^j_{\ \nu}b^k_{\ \rho} \right] = 0 \\
\frac{\delta S}{\delta \omega^i_{\ \mu}} &= \epsilon^{\mu\nu\rho} \left[ \alpha_3\, R_{i\nu\rho} + a\, T_{i\nu\rho} + \alpha_4\, \epsilon_{ijk}b^j_{\ \nu}b^k_{\ \rho} \right] = 0.
\end{aligned}
\end{align}
The constraints in the above model are \cite{Blagojevic:2004hj} listed in Table \ref{Tab:Constraints MB}
\begin{table}[t]
\caption{\label{Tab:Constraints MB}Constraints of the theory.}
\begin{center}
\begin{tabular}{l c c}
\br
& First Class $\psi$ & Second class $\chi$ \\
\mr
Primary & $\phi_i^{\ 0}\;, \Phi_i^{\ 0}$ & $\phi_i^{\ \alpha}$, $\Phi_i^{\ \alpha} $ \\[0.4ex]
Secondary & $\bar{\mathcal{H}}_i\,, \bar{\mathcal{K}}_i$ &  \\
\br
\end{tabular}
\end{center}
\end{table}
with the definitions
\begin{align}
\label{Rel qtys MB}
\begin{aligned}
\phi_i^{\ \mu} &= \pi_i^{\ \mu} - \alpha_4\,\epsilon^{0\alpha\beta}\, b_{i\beta}\,\delta^\mu_\alpha \\
\Phi_i^{\ \mu} &= \Pi_i^{\ \mu} - \epsilon^{0\alpha\beta} \left( 2 a\, b_{i\beta} + \alpha_3\, \omega_{i\beta} \right) \delta^\mu_\alpha\\
\bar{\mathcal{H}}_i &= - \left[ \epsilon^{0\alpha\beta}\!\left( a\, R_{i\alpha\beta} + \alpha_4\, T_{i\alpha\beta} - \Lambda \epsilon_{ijk} b^j_{\ \alpha} b^k_{\ \beta} \right)\right] - \nabla_{\!\alpha} \phi_i^{\ \alpha} + \epsilon_{ijk}\, b^j_{\ \alpha} \left( p\,\phi^{k\alpha} + q\, \Phi^{k\alpha} \right) \\
\bar{\mathcal{K}}_i &= -\left[ \epsilon^{0\alpha\beta} \left( a\,T_{i\alpha\beta} + \alpha_3\,R_{i\alpha\beta} + \alpha_4\,\epsilon_{ijk} b^j_{\ \alpha} b^k_{\ \beta} \right) \right] - \nabla_{\!\alpha} \Phi_i^{\ \alpha} - \epsilon_{ijk}\, b^j_{\ \alpha} \phi^{k\alpha}\\
p &= \frac{\alpha_3\Lambda + \alpha_4a}{\alpha_3 \alpha_4 - a^2}\,;\qquad q = -\frac{\alpha_4^2 + a\Lambda}{\alpha_3\alpha_4 - a^2}.
\end{aligned}
\end{align}
The gauge generator $G$ which generates gauge symmetries is, by the Dirac principle, a linear combination of all first class constraints $$G= \varepsilon_A \psi_A\,,$$ the sum in $A$ running over the complete first-class sector. The symmetries of the basic variables are then obtained through $\delta q = \lbrace q, G \rbrace^* = \varepsilon_A \lbrace q, \psi_A \rbrace^*$ where $\lbrace\,,\,\rbrace^*$ indicates Dirac brackets obtained through elimination of the second class sector $\chi$ \cite{Banerjee:2009vf}.

Now, we know that all the gauge parameters $\varepsilon_A$ are not independent, as the number of independent gauge parameters must equal the number of independent primary first-class constraints. In the off-shell method that we adopt \cite{Henneaux:1990au,Banerjee:1999yc,Banerjee:1999hu}, the restriction on $\varepsilon_A$ is obtained from the commutativity of gauge variations and time evolution $\delta \bullet \frac{d}{dt} \equiv \frac{d}{dt} \bullet \delta\,.$  Explicit calculations \cite{Banerjee:2009vf} yield the generator
\begin{align}
\label{Gen MB}
\begin{aligned}
G=\int \textrm{d}^2x ~\left(\dot{\tau}^i\,\pi_i^{\ 0} + \tau^i\left[\bar{\mathcal{H}_i}- \varepsilon_{ijk} \big( \omega^j_{\ 0} - p\,b^j_{\ 0}\big)\pi^{k0} + q \,\varepsilon_{ijk}\,b^j_{\ 0}\Pi^{k0} \right]\right. \\
\left.+ \dot{\sigma}^i\Pi_i^{\ 0} + \sigma^i\left[\bar{\mathcal{K}_i}-\varepsilon_{ijk}\big(b^j_{\ 0}\,\pi^{k0} + \omega^j_{\ 0}\,\Pi^{k0}\big)\right]\right)
\end{aligned}
\end{align}
where $\tau^i$ and $\sigma^i$ are six independent gauge parameters. We are now ready to study the gauge symmetries of the model utilising this generator.

\section{Hamiltonian and Poincare gauge: role of trivial symmetries}

The generator \eqref{Gen MB} yields the following symmetries of the basic fields
\begin{align}
\label{symm MB}
\begin{aligned}
\delta_H b^i_{\ \mu} &= \nabla_\mu\tau^i - p \,\epsilon^i_{\ jk} \,b^j_{\ \mu} \tau^k + \epsilon^i_{\ jk}\,b^j_{\ \mu} \sigma^k ,\\
\delta_H \omega^i_{\ \mu} &= \nabla_\mu \sigma^i - q \,\epsilon^i_{\ jk} \,b^j_{\ \mu} \tau^k.\\
\end{aligned}
\end{align}
A first comparison of the two sets of symmetries \eqref{PGT deltas} and \eqref{symm MB} reveals that the Hamiltonian symmetries are explicitly dependant on the coupling parameters of the various terms in the action \eqref{action TMG}, in contrast to the PGT symmetries. Also, we note that the gauge parameters of the two symmetries are different. So we need to map the gauge parameters of \eqref{symm MB} into the PGT gauge parameters. This is achieved through the field dependent map\footnote{For a construction of the map, using Noether identities corresponding to the two sets of symmetries, see \cite{Banerjee:2011cu}.} \cite{Blagojevic:2004hj}
\begin{align}
\label{map}
\tau^i = -\xi^\rho\,b^i_{\ \rho} \qquad\&\qquad \sigma^i = -\theta^i - \xi^\rho \omega^i_{\ \rho}\,.
\end{align}
Using this map, the Hamiltonian symmetries can be written in the form $$\delta_H \sim \delta_{\scriptscriptstyle PGT} + \text{equations of motion}\,.$$ It may seem that the two symmetries are thus equivalent only on-shell. To show that this is not the case, we write out their form explicitly
\begin{align}
\label{mapped MB}
\begin{aligned}
\delta_H b^i_{\ \mu} &= \delta_{\scriptscriptstyle PGT} b^i_{\ \mu} + \frac{\alpha_3}{2(\alpha_3\alpha_4-a^2)}\,\eta^{ij}\,\xi^\rho\,\epsilon_{\mu\nu\rho} \,\frac{\delta S}{\delta b^j_{\ \nu}} - \frac{a}{2(\alpha_3\alpha_4-a^2)}\,\eta^{ij}\,\xi^\rho\,\epsilon_{\mu\nu\rho} \,\frac{\delta S}{\delta \omega^j_{\ \nu}} \\
\delta_H \omega^i_{\ \mu} &= \delta_{\scriptscriptstyle PGT} \omega^i_{\ \mu} - \frac{a}{2(\alpha_3\alpha_4-a^2)}\,\eta^{ij}\,\xi^\rho\,\epsilon_{\mu\nu\rho} \,\frac{\delta S}{\delta b^j_{\ \nu}} + \frac{\alpha_4}{2(\alpha_3\alpha_4-a^2)}\,\eta^{ij}\,\xi^\rho\,\epsilon_{\mu\nu\rho} \,\frac{\delta S}{\delta \omega^j_{\ \nu}} \\
\end{aligned}
\end{align}
To cast the balance terms of $\delta \equiv \delta_H-\delta_{\scriptscriptstyle PGT}$ into the form \eqref{trivial gauge gen}, we first write down the appropriate $\Lambda$ matrix:
\begin{align}
\label{Lambda}
\begin{aligned}
\Lambda_{\left( b^i_{\ \mu},\, b^j_{\ \nu} \right)} &= \frac{\alpha_3}{2(\alpha_3\alpha_4 - a^2)}\,\eta^{ij}\,\xi^\rho \epsilon_{\mu\nu\rho} \qquad &
	\Lambda_{\left( b^i_{\ \mu},\,\omega^j_{\ \nu} \right)} &= \frac{-a}{2(\alpha_3\alpha_4 - a^2)}\,\eta^{ij}\,\xi^\rho \epsilon_{\mu\nu\rho} \\
\Lambda_{\left( \omega^i_{\ \mu},\, b^j_{\ \nu} \right)} &= \frac{-a}{2(\alpha_3\alpha_4 - a^2)}\,\eta^{ij}\,\xi^\rho \epsilon_{\mu\nu\rho} \qquad &
	\Lambda_{\left( \omega^i_{\ \mu},\, \omega^j_{\ \nu} \right)} &= \frac{\alpha_4}{2(\alpha_3\alpha_4 - a^2)}\,\eta^{ij}\,\xi^\rho \epsilon_{\mu\nu\rho}\,.
\end{aligned}
\end{align}
The anti-symmetry of the above structure can now be easily demonstrated. As an example,
\begin{align*}
\Lambda_{\left( b^i_{\ \mu},\, \omega^j_{\ \nu} \right)} &= \frac{-a}{2(\alpha_3\alpha_4 - a^2)}\,\eta^{ij}\,\xi^\rho \epsilon_{\mu\nu\rho}  \nonumber\\
&= - \frac{-a}{2(\alpha_3\alpha_4 - a^2)}\,\eta^{ji}\,\xi^\rho \epsilon_{\nu\mu\rho} = - \Lambda_{\left( \omega^j_{\ \nu},\, b^i_{\ \mu} \right)}\,.
\end{align*}

So, we see that indeed `$\delta$' is of the required form \eqref{trivial gauge gen} for trivial gauge symmetries. Thus the invariance of the action can now be achieved under these `balance' symmetries $\delta \equiv \delta_H - \delta_{\scriptscriptstyle PGT}$ without the use of any equations of motion. Hence we see, that from a canonical point of view, $$ \delta_H \equiv \delta_{\scriptscriptstyle PGT}$$ as their difference is just an un-physical `trivial' symmetry.


\end{document}